\documentclass[fleqn,twoside]{article}
\usepackage{espcrc2}
\usepackage{epsfig}

\usepackage{graphicx}
\usepackage[figuresright]{rotating}

\def\bea{\begin{eqnarray}}
\def\eea{\end{eqnarray}}

\newcommand{\AmS}{{\protect\the\textfont2
  A\kern-.1667em\lower.5ex\hbox{M}\kern-.125emS}}

\title{
\vspace{-3.0cm}
       \rightline{\normalsize CERN-TH/2001-288}
         \rightline{\normalsize IFIC/01-59, FTUV/011030}
       \vspace{2.0cm}
Ginsparg-Wilson fermions: practical aspects and applications}

\author{P.~Hern\'andez \address[]{Theory Division, CERN,\\
 1211 Geneva 23, Switzerland} 
 \thanks{On leave from Dpto. F\'{\i}sica Te\'orica, Universidad de Valencia.}}
       
\begin{document}

\begin{abstract}
I review recent progress in the implementation of lattice fermions satisfying the Ginsparg--Wilson relation and some physical applications. 
\vspace{1pc}
\end{abstract}

\maketitle

\section{Introduction}

It has been established that lattice Dirac operators $D$ can be constructed which are local, do not suffer from the doubling problem and satisfy the Ginsparg--Wilson (GW) relation \cite{gwold}:
\bea
\{ D, \gamma_5 \} = D \gamma_5  D \;\;\leftrightarrow \;\;\{D^{-1} , \gamma_5\} = \gamma_5.
\label{eq:gw}
\eea 
This relation ensures that Ward identities associated with a standard chiral
transformation are satisfied on shell at finite $a$ \cite{gwold,gwnew}. 
Furthermore, this relation implies an exact lattice chiral symmetry \cite{exsym} and index theorem \cite{hln,exsym}.

Two apparently disconnected avenues led to operators 
satisfying the GW relation. The first one was the 
lattice version of the five dimensional (5D) Dirac fermion Yukawa-coupled to a domain wall (DW), which naturally leads to a chiral fermion in four dimensions, attached 
to the defect. On a lattice such a construction involves 
in general two lattice spacings $a, a_s$ (in the physical four and the extra dimensions respectively). If the number of sites in the extra dimension, $N_s$,  
is infinite, one still expects some kind of discretized chiral fermion by analogy with the continuum analysis \cite{dwlat}. Indeed it can be shown that in this limit, at finite lattice spacing,  
the effective action describing the light fermionic degrees of freedom is that of a 4D GW fermion \cite{dwgw}.
 For the standard DW construction of \cite{dwlat}, the 4D Dirac operator describing the light degrees of freedom at finite $N_s$ is 
\bea
a D_{N_s} = 1- \gamma_5 \frac{(1+\bar{Q})^{N_s} - (1-\bar{Q})^{N_s}}{(1+\bar{Q})^{N_s} + (1-\bar{Q})^{N_s}}, \nonumber\\
\bar{Q} \equiv  \gamma_5 \frac{a_s A}{2 - a_s A}\;\;\;\;A \equiv m_0 - D_W.
\label{eq:dwdn}
\eea
$m_0$ is the height of the domain wall and $D_W$ is the standard Wilson Dirac operator. Formally
\bea
\lim_{N_s\rightarrow \infty} a D_{N_s} \rightarrow 1 -\gamma_5 \rm{sign}(\bar{Q})  
= 1 - \bar{Q}/\sqrt{\bar{Q}^2},
\label{sqrt} 
\eea
which satisfies the GW relation. 
If also the limit $a_s \rightarrow 0$ is taken, the overlap operator \cite{overlap} is obtained\footnote{Note that 
if $m_0 a\neq 1$, the operator of eq.~(\ref{eq:gw}) is not properly normalized. For simplicity, the formulae below assume $m_0 a = 1$ but are easily generalized.}: 
\bea
\lim_{a_s\rightarrow 0} \lim_{N_s\rightarrow \infty} a D_{N_s} \rightarrow 1 -\gamma_5 \rm{sign}(Q)\;\;,Q=\gamma_5 A.
\label{eq:ov}
\eea 
The locality of the operators of eqs.~(\ref{sqrt}) and (\ref{eq:ov}) was 
shown in \cite{locov,locdw}. 

The second avenue was the search for lattice Dirac operators obtained from a renormalization group (RG) 
blocking of the continuum one \cite{gwold}. 
Starting from a chirally symmetric Dirac operator in 
the continuum, the effective operator on the lattice is not chirally symmetric in general, but 
retains a remnant of the original symmetry encoded in the GW relation.
A fixed-point (FP) Dirac operator \cite{fp} is invariant under a RG transformation and, thus, not only does it satisfy the GW relation \cite{gwnew}, but it is  
in principle free of discretization 
errors. In practice, however, such an operator is very difficult to find and no
explicit construction exists.  

In this review I concentrate on the uses of GW fermions for lattice QCD. 
For a review on other topics including chiral gauge theories see \cite{kiku}.
  
\section{Chiral symmetry at finite $a$}

Given that it is very expensive to simulate GW fermions, it is important to assess the need for 
 an exact chiral symmetry at finite $a$ in lattice QCD. There are at least three very good 
reasons to preserve such a symmetry.

{\it 1.} The renormalization of fermion operators is enormously simplified 
by the exact chiral symmetry. For any Dirac operator $D$ satisfying 
the GW relation of eq.~(\ref{eq:gw}), we can define the chiral projections
\bea
\Psi_{R,L} = {\hat P}_{\pm} \Psi,\;\;\;\; \bar{\Psi}_{L,R} = \bar{\Psi} P_{\pm}
\label{chicomp}
\eea
with $P_{\pm} \equiv (1\pm \gamma_5)/2$ and ${\hat P}_{\pm} \equiv (1\pm \gamma_5 (1- a D))/2$, so that the action density splits as
\bea
\bar{\Psi} D \Psi = \bar{\Psi}_L D \Psi_L + \bar{\Psi}_R D \Psi_R,  
\label{action}
\eea
implying an exact $SU(N_f)_L \times SU(N_f)_R$ flavour symmetry under which 
the chiral components transform in the standard way:
\bea
\Psi_L \rightarrow V_L \Psi_L, \;\;\;\;\; \Psi_R \rightarrow V_R \Psi_R \;\;\; V_{L,R} \in SU(N_f)_{L,R}.\nonumber
\eea
Quark masses should be included by adding $\bar{\Psi}_L m \Psi_R + \bar{\Psi}_R m^\dagger \Psi_L$ to the action. The usual spurion analysis implies that operators can mix only if they have the same transformation 
properties under the exact lattice symmetry
\bea
\Psi_L \rightarrow V_L \Psi_L \;\;\;\;\; \Psi_R \rightarrow V_R \Psi_R\;\;\;\;\; m \rightarrow V_L m V_R^\dagger. 
\label{spurion}
\eea
Several important results follow in a straightforward way from this analysis:

$\bullet$ there is a conserved axial current and no additive mass 
renormalization; 

$\bullet$ two and four-fermion operators relevant to weak matrix elements 
can be constructed (using these chiral 
components) in chiral representations of the symmetry group of eq.~(\ref{spurion}) 
in such a way that their mixing is the same as in any chirally symmetric continuum regularization.  

Recently the perturbative renormalization constants for quark bilinears and four-fermion operators have been computed for both DW and overlap fermions \cite{aoki,al}.

{\it 2.} $O(a)$ improvement is almost automatic. A similar spurion analysis as 
that outlined above shows that there are no invariant operators of dimension $d=5$ 
 so the action is $O(a)$-improved off shell. Similarly the quark bilinears are easily $O(a)$-improved if they are constructed in the same chiral representations as their continuum counterparts \cite{impro}.

{\it 3.} The exact chiral symmetry makes it possible to 
explore the regime of the light quark masses. Naively one would think that
this is anyway very difficult since the pion Compton wavelength would not fit in the lattice volumes 
accessible to simulations: it would not be possible to ensure the 
condition $M_\pi L \gg 1$ resulting in 
large finite volume effects. Fortunately (quenched) chiral perturbation theory (Q)$\chi$PT predicts these effects \cite{gl,damgaard}, including those of 
topological zero modes, and a finite size scaling analysis in the regime
$M_\pi L \leq 1$, is actually a very useful tool to extract the physical low energy constants. 
Recently this technique has been applied to quenched QCD using GW fermions \cite{fss,fss2,degrand,bern} to extract the low energy constant $\Sigma$.  

 
\section{Implementation of GW fermions}

{\it Overlap fermions}

The difficulty in the implementation of overlap fermions  is in constructing 
the square root or sign function of eq.~(\ref{eq:ov}). Several methods 
have been proposed to approximate these functions, which are variants of two basic approaches:

$\bullet$ Polynomial approximations (PA), e.g. Chebyshev or Legendre \cite{fss2,pa}: 

\begin{flushleft} $D_N \simeq 1- \gamma_5 Q P_N(Q^2)$ \end{flushleft}

$\bullet$ Rational approximations (RA) \cite{ra,low}: 
\bea
D_N  \simeq 1-\gamma_5 Q \left(c_0 + \sum_{k=1}^N \frac{c_k}{Q^2 + q_k}\right).
\eea
In this case, the $N$ inversions $(Q^2+ q_k)^{-1}$ can all be obtained in 
one go, using a multimass solver conjugate gradient (CG) \cite{mmcg}. The nested
CG necessary to invert the resulting operator can be avoided by rewriting the RA as a truncated continued 
fraction, which in turn can be rewritten as a Gaussian path integral with extra 
fermionic fields \cite{cf}. The problem reduces to inverting a 
local 5D operator very similar to the DW.   

\begin{figure}[htb]
\vspace{-.5cm}
\epsfig{file=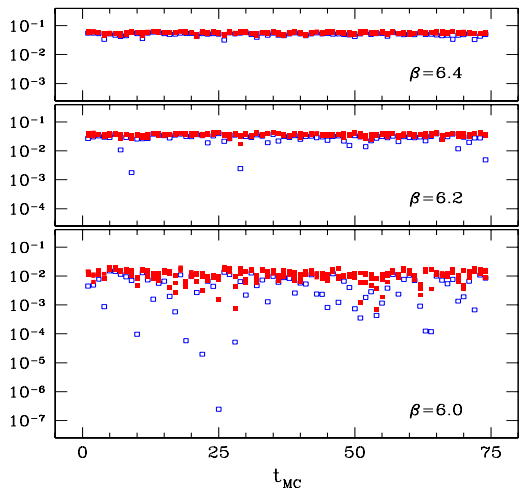,width=7.cm, height=6.5cm}
\vspace{-1.cm}
\caption{MC evolution of the four lowest eigenvalues of $Q^2$ \cite{locov}.}
\label{fig:spec}
\vspace{-.8cm}
\end{figure}

Naively, the cost increase with respect to Wilson fermions is the number, $n$,
of additional $Q$ matrix$\times$vector multiplications that are 
needed to approximate the square root: $2 N$ for the PA, and twice the number of CG iterations needed to invert the factor $Q^2+q_1$, with the minimum pole, for RA. 
Asymptotically, the convergence of the approximation is exponential in $n$ in both cases:
\bea
||D_N - D_{ov} || < e^{-  \sqrt{\epsilon} n},\;\epsilon = \lambda_{min}(Q^2)/\lambda_{max}(Q^2),\nonumber
\eea
but it is controlled by the condition number of the 
matrix $Q^2$. 
Figure~\ref{fig:spec} shows the distribution of the four lowest eigenvalues
of $Q^2$ for three values of $\beta$. At large $\beta$ the spectrum shows a gap,  
which shrinks with lowering $\beta$ and becomes populated by isolated eigenvalues that can be orders of magnitude smaller. The 
presence of these eigenvalues results in a very poor convergence of any 
approximation to the sign function. 

{\it DW fermions}

The cost of DW fermions is less obvious. First, there is 
a factor of $N_s$ more computations involved owing to the extra dimension. 
The asymptotic convergence to the limit $N_s \rightarrow \infty$ is very similar to the convergence in $n$ of the approximations to 
the overlap: it is exponential in $N_s$, with a rate determined by the lowest eigenvalue of the operator $\bar{Q}^2$ of eq.~(\ref{eq:dwdn}). 
 The distribution of the low-lying eigenvalues of $\bar{Q}^2$ is very similar
to that of $Q^2$ in Fig.~\ref{fig:spec} \cite{ringberg}, implying that 
the convergence in $N_s$ also becomes extremely slow at accessible $\beta$'s. 
It is customary to quantify the 
deviations from the $N_s = \infty$ limit by the anomalous term in the axial 
Ward identity properly normalized, i.e. the so-called residual mass. The dependence of the residual mass on $N_s$ has been extensively studied numerically \cite{rbc,cppacs}, and indeed it is found that its exponential 
decrease with $N_s$ is very slow. A recent study showed that the small
eigenvalues of $Q^2$ are responsible for 
the residual mass at large $N_s$ \cite{taniguchi}.  
An additional cost of DW fermions comes from the fact that the number of CG 
iterations needed to invert the 5D operator is considerably larger than for 
the 4D Wilson operator.  
Although a more detailed comparison is needed, there is at present no evidence that the DW formulation is more efficient than the overlap at some fixed 
accuracy of the GW relation \cite{nn,j}. 

{\it FP Dirac operator} 

\begin{figure}[ht]
\vspace{-.4cm}
\begin{center}
\epsfig{file=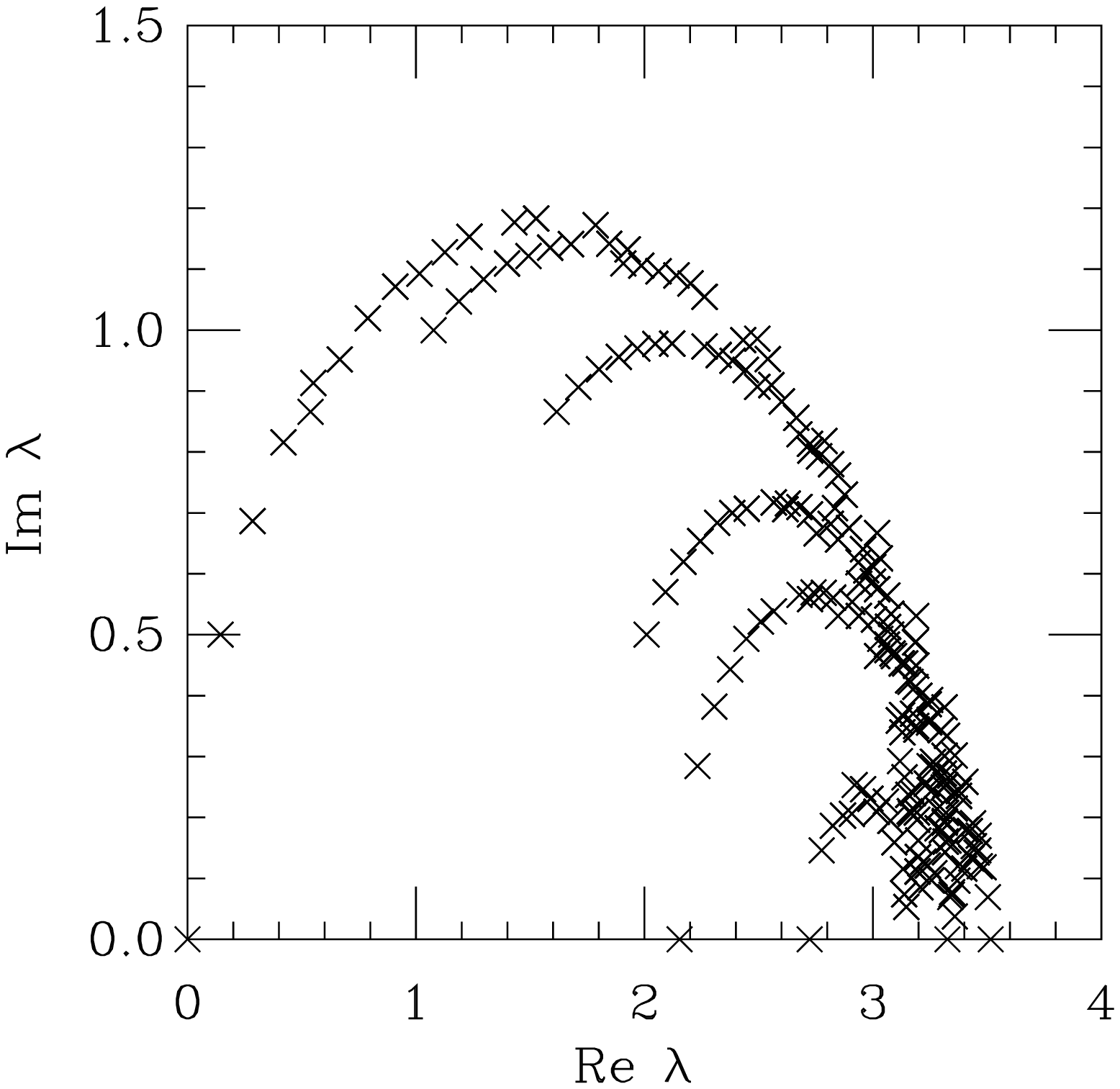,height=3cm,width=3.5cm}  \epsfig{file=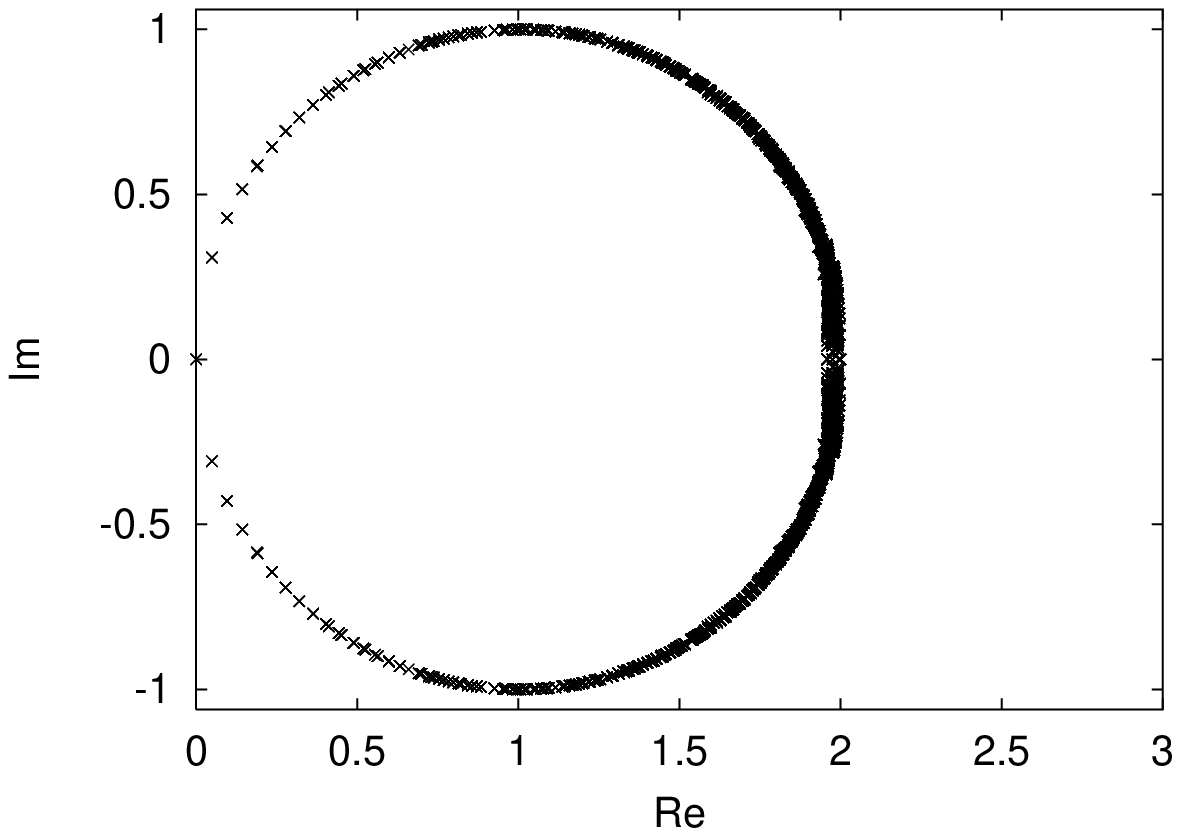,height=3cm,width=3.5cm} \\
\epsfig{file=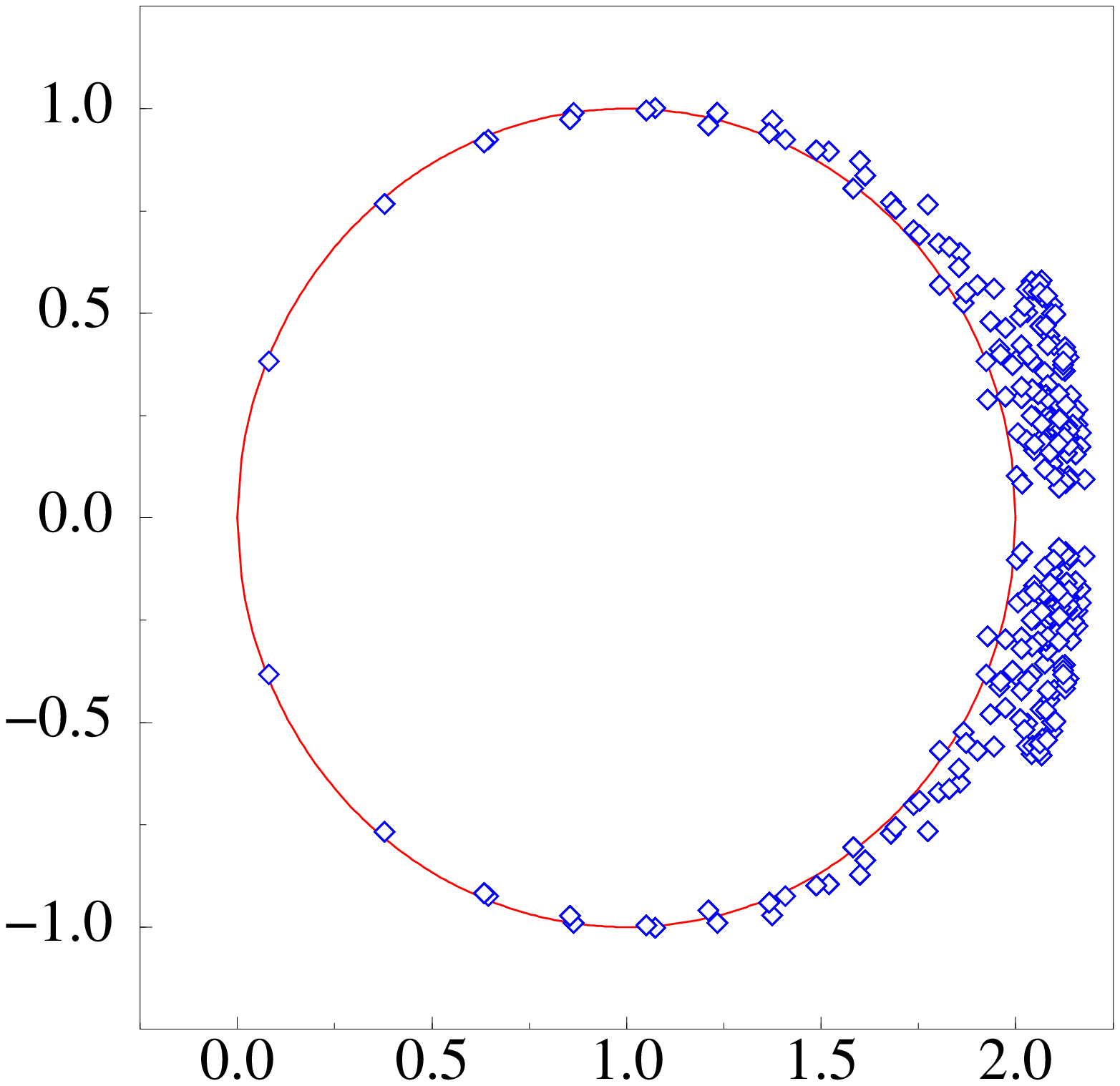,height=3cm,width=3cm}  \epsfig{file=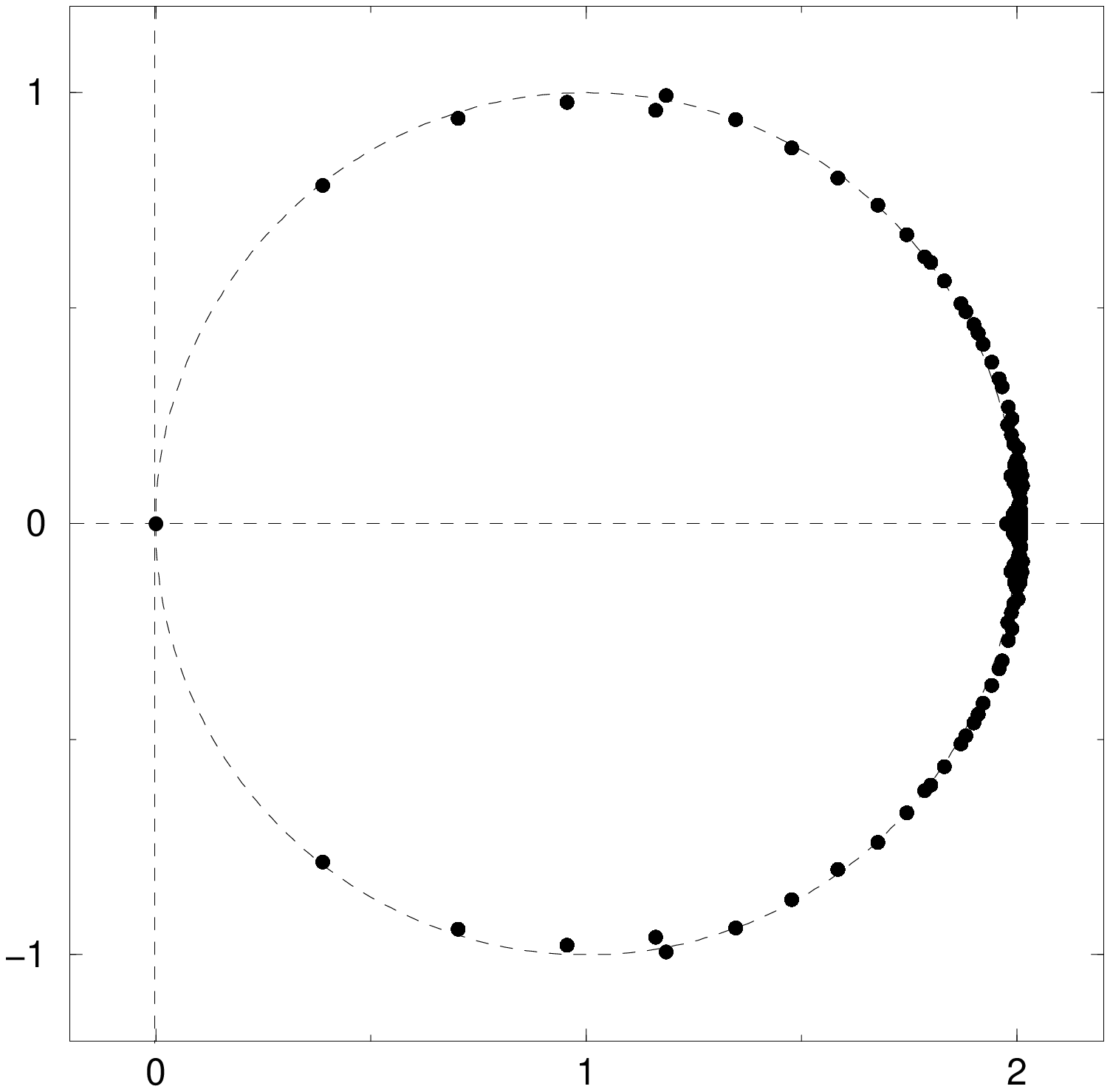,height=3cm,width=3cm} 
\end{center}
\vspace{-1cm}
\caption{Free spectrum of the actions of refs. \cite{degrand,fp1} (up) and \cite{gw2,bern} (down).}
\vspace{-.5cm}
\label{fig:fp}
\end{figure}

In recent years  a big effort has been devoted to find, among the Dirac operators with 
all the allowed hypercubic couplings, the one that approximates the closest either the FP operator or simply the GW relation. Fig.~\ref{fig:fp} shows the free spectrum of four such operators presented 
in refs.~\cite{degrand,fp1,gw2,bern} (ordered from left to right and up to down).
The spectrum 
should lie on a circle if the approximation were perfect. These operators are also considerably more costly than Wilson, e.g. a factor $\sim$ 20 for  
an accuracy of the GW relation of about $1\%$ \cite{bern}. 

\section{Improvements}

As we have seen, the cost of the implementation of overlap or 
DW fermions is inversely proportional to the square root of the gap 
of the operators $Q^2$ or $\bar{Q}^2$ respectively. Several ideas to improve
the convergence in this respect have been proposed and studied:

$\bullet$ {\em A small eigenspace of $O(10)$ eigenvalues is treated exactly in the
limit $n, N_s \rightarrow \infty$.} 

 In the context of the overlap this means that the exact operator 
is computed in this eigenspace, while a PA or RA is used only in the space orthogonal to it \cite{low}. The convergence is then controlled by the smallest eigenvalue of the orthogonal subspace, which can 
be orders of magnitude larger than the smallest eigenvalue of $Q^2$.
Analogously, in the case of the DW operator
the challenge is to construct a  local 5D operator such that its
effective 4D action is  ${\tilde D}_{N_s} = D_{N_s} + \Delta$, where 
$D_{N_s}$ is the operator in eq.~(\ref{eq:dwdn}), $\bar{Q} |\Phi_i\rangle = \mu_i |\Phi_i\rangle$  and
\bea
a \Delta \equiv
-\gamma_5 \sum_{i=1}^p \left({\rm sign}(\mu_i) - \frac{\alpha^+_i - \alpha_i^-}{\alpha_i^+ + \alpha_i^-} \right) |\Phi_i\rangle\langle\Phi_i| \nonumber
\eea
with $\alpha_i^{\pm} = 1\pm \mu_i$. The convergence in $N_s$ of ${\tilde D}_{N_s}$ is controlled by $\epsilon = |\mu_{p+1}| \gg 
|\mu_1|$. 
There is no unique 5D operator satisfying this criterion. At least two explicit methods have been described in refs. \cite{eh,ringberg}.
Note that the condition number of the 5D operator might depend on the implementation.

$\bullet$ {\em The gap of $Q^2$ and $\bar{Q}^2$ is increased by improving the gauge action.} \cite{cppacs}

\begin{figure}[htb]
\vspace{-.8cm}
\epsfig{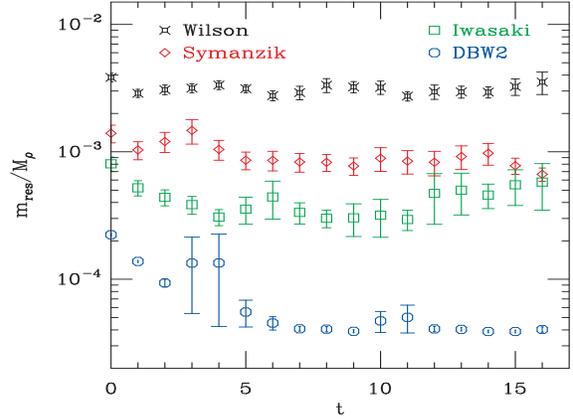}
\vspace{-1.5cm}
\caption{Plateau of the residual mass over $M_\rho$ at $a^{-1}\simeq 2$ GeV and $N_s=16$ \cite{comp}.}
\vspace{-.8cm}
\label{fig:comp}
\end{figure}

This is a cheap solution in principle. A comparison of different 
actions in the DW construction has been carried out in \cite{comp}. 
Fig.~\ref{fig:comp} 
shows the time dependence of the residual mass for 
the Wilson action, the one-loop and tadpole-improved Symanzik action and two
RG-improved actions: Iwasaki's and DBW2 \cite{dbw2}. The improvement 
of the DBW2 action is remarkable. Some understanding of this effect comes 
from the 
fact that this action strongly supresses small (and not so small) instantons \cite{marga}. The overimprovement of the DBW2 action in particular is very large, resulting in a very slow thermalization of the topological charge \cite{comp}. 
It will be important to clarify  
if the use of these gauge actions results in a significant improvement also 
when the method of treating exactly a small eigenspace is implemented.

$\bullet$ {\em Substitute $D_W$ by an improved operator in the overlap or DW constructions.} 

If $D_W$ in eq.~(\ref{eq:dwdn}) is substituted by an operator that is closer 
to satisfy the GW relation, the approximation of the overlap operator or 
the limit $N_s\rightarrow \infty$ of the DW are much faster,  
since the size of the gap of the corresponding $Q^2$ (or $\bar{Q}^2$) is close
to $1$ \cite{bietenholz}. This idea has been implemented in the overlap context with the operators of Fig.~\ref{fig:fp}. The results show that in general 
the gain in the convergence of the approximations of the sign function does
not quite compensate the overhead of using these actions. On the other hand, the overlap operator is more local (at least in those cases where fat links are  
not involved) and also the scaling properties are improved in the case of the truncated
FP actions (see results in \cite{fp1,bern}).    

Finally, there is room for algorithmic improvement. 
The authors of \cite{lippert} pointed out that, in the context of the RA, the vector updates in the multimass CG solver are costly 
for a large number of poles. They proposed a new RA
which requires less poles than previously used ones and from which 
they can derive rigorous error bounds. 
The authors of \cite{borici} proposed a strategy to 
search for truncated continued fractions that result in better conditioned
5D operators.

\section{A few physics results}

In the past year a number of QCD simulations with GW fermions have been performed. New results on the weak matrix elements using standard DW fermions have been presented. 
I refer to \cite{guido} for a description of these results. 
An important step forward this year has been the demonstration that simulations with overlap fermions are also feasible in the quenched approximation. Results on hadron spectroscopy and
low energy constants have been presented by several groups  using 
both the standard overlap operator \cite{ghr,hjlw2} and also improved ones \cite{degrand,bern,dong}. 

\begin{figure}[htb]
\vspace{-1.cm}
\epsfig{file=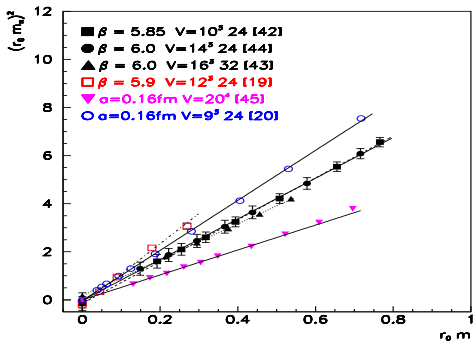,width=7.cm,height=5cm}
\vspace{-1.cm}
\caption{$m_\pi^2$ vs. bare $m$ in units of $r_o$.}
\vspace{-.8cm}
\label{fig:mpi}
\end{figure}

\begin{figure}[htb]
\vspace{-1.2cm}
\epsfig{file=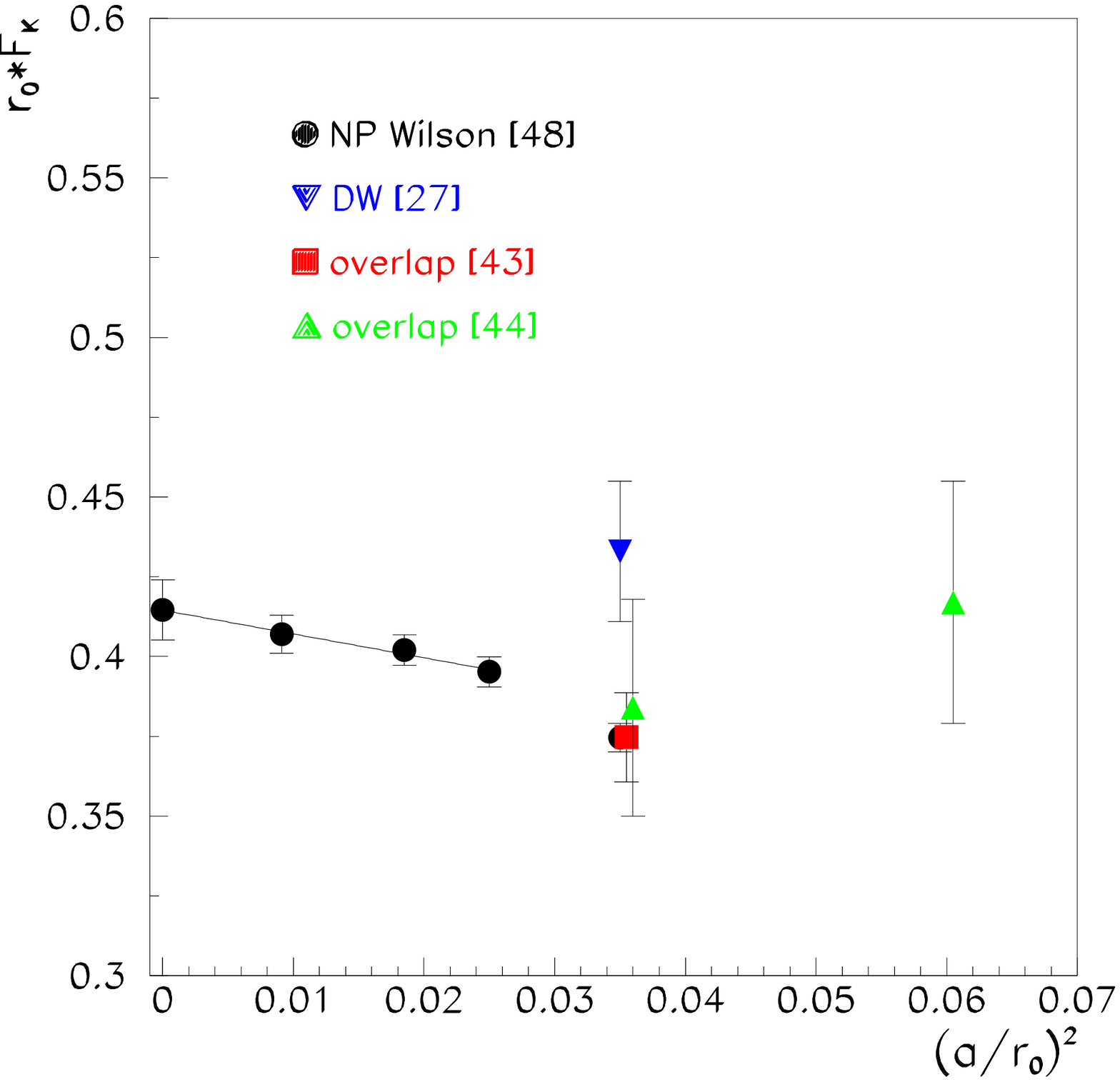,width=7.cm,height=5cm}\\
\epsfig{file=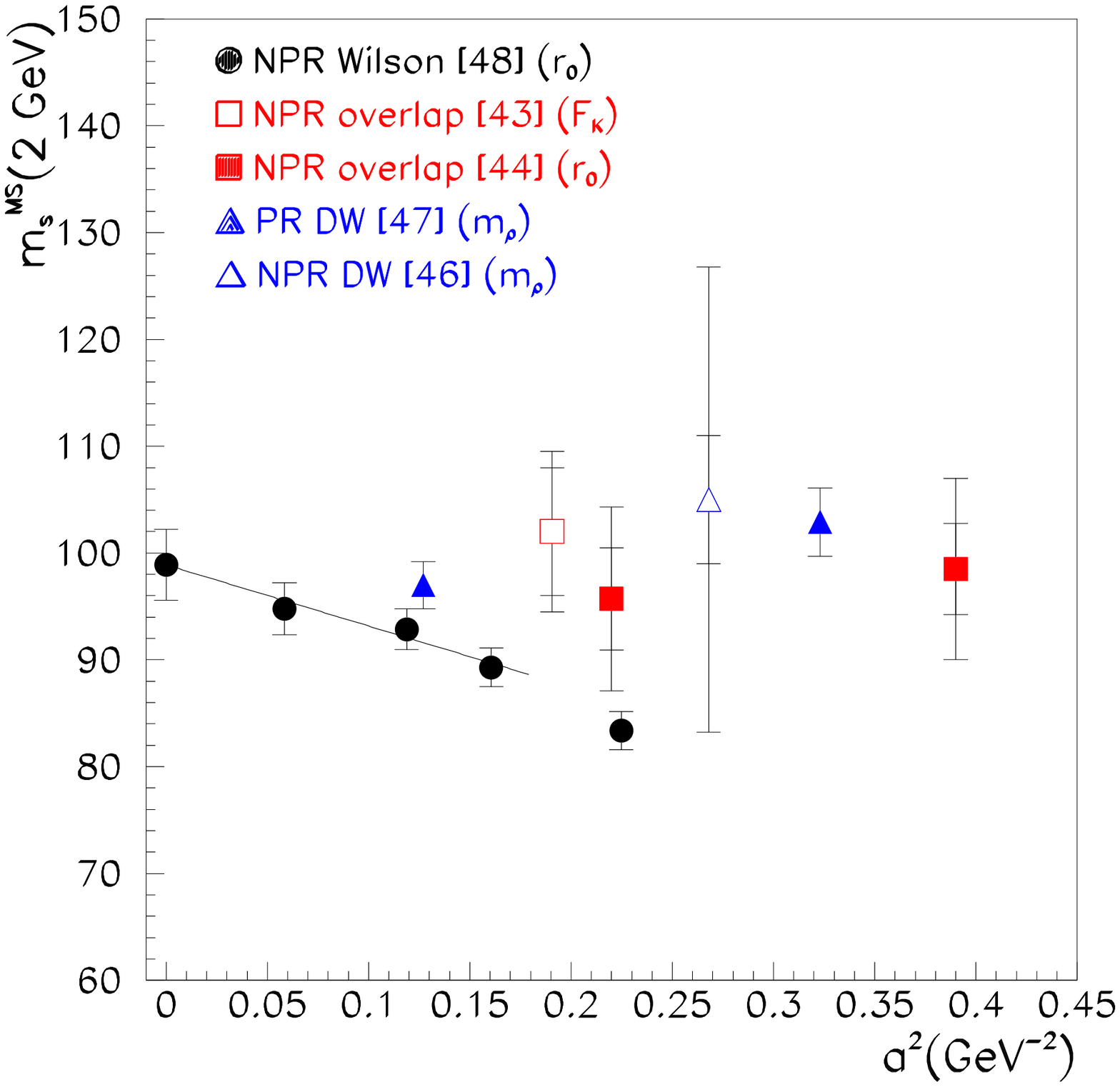,width=7.cm,height=5cm}
\vspace{-1cm}
\caption{$a$ scaling of $F_K$ and $m_s^{\overline{MS}}$(2GeV). 
In the lower plot $a$ is set by different inputs as indicated. The first 
error bar is statistical, the second includes the error on the renormalization factor.}
\vspace{-.5cm}
\label{fig:mpifk}
\end{figure}

Pseudoscalar and scalar two-point functions have been computed at $a = 0.09-0.16 fm$ in various physical volumes. 
Fig.~\ref{fig:mpi} shows the pion mass square versus the bare quark mass 
in units of $r_0$ extracted from the pseudoscalar correlator (full symbols)
or the pseudoscalar minus the scalar one (open symbols) \footnote{This combination is expected to be less sensitive to topological zero modes for small 
quark masses\cite{rbc}.}. 
Within errors the behaviour is linear and
consistent with a vanishing intercept, as expected from the exact chiral symmetry of these actions. Assuming lowest order
Q$\chi$PT, the low energy constant $\Sigma$ can be extracted from the slope. 
The needed renormalization factor, $Z_m^{-1} = Z_S = Z_P$, has 
been obtained non-perturbatively \cite{hjlw,ghr}.
This same renormalization factor can be used to obtain the renormalized
 $\Sigma$ from a finite-size scaling analysis in a completely different regime \cite{fss2,degrand,bern}. The agreement of these two quantities is very good \cite{hjlw2,bern}. Also a value for the 
 strange quark mass and decay constant, $F_K$, can be obtained from these measurements. Note that the latter needs no renormalization in GW regularizations. 
Figs.~\ref{fig:mpifk}
show the comparison of the overlap results for these quantities   
with earlier ones obtained with DW fermions \cite{rbc,dw1,dw2} and $O(a)$-improved Wilson fermions \cite{wilson}. These results show no surprises and, although the statistical errors are 
still large, they seem to indicate rather small $O(a^2)$ discretization errors
for overlap fermions. 

\section{Conclusions}

Having an exact chiral symmetry at finite $a$ in lattice QCD permits the 
exploration of the regime of light quark masses and might turn out to be 
essential in solving long-standing problems, such as the $\Delta I =1/2$ rule 
or the prediction of $\epsilon'/\epsilon$. Unfortunately, it is also 
very costly and a simple rule of thumb gives about 2 orders of magnitude
overhead with respect to Wilson fermions. Several ideas to improve the 
situation have been proposed and implemented, but they have not yet 
resulted in a significant cost improvement in logarithmic terms. In spite
of this situation, several quenched simulations have been successfully 
carried out with GW fermions and show that these actions 
do not have large discretization errors of $O(a^2)$ and can  
reach a regime of light quark masses  which is not accessible to Wilson 
fermions. 


\end{document}